\documentclass[a4paper,12pt,nofootinbib]{revtex4}
\usepackage{amssymb}
\usepackage{epsfig}

\begin{document}


\pagenumbering{arabic}

\flushbottom


\title{THE STAR-FORMATION EFFICIENCY AND DENSITY OF THE DISKS OF
SPIRAL GALAXIES}

\affiliation{Sternberg Astronomical Institute, Universitetskii pr.
13, Moscow, 119992 Russia}

\author{\firstname{A.~V.}~\surname{Zasov}}
\email[E-mail: ]{zasov@sai.msu.ru}\noaffiliation

\author{\firstname{O.~V.}~\surname{Abramova}}
\noaffiliation

\received{March 10, 2006}
\revised{May 14, 2006}

\begin{abstract}
For four well studied spiral galaxies (M33, M81, M100 and M101) we
consider the dependencies of star formation rate ($SFR$) and star
formation efficiency ($SFE = SFR/M_{gas}$) both on the radial
distance $R$ and on some kinematic parameters of galactic discs.
To estimate $SFR(R)$ we used a combined $UV+FIR$ method based on
the UV profiles corrected for the interstellar extinction
presented by Boissier et al~\cite{main}. It is demonstrated that
the most tight correlation, similar for all galaxies we
considered, exists between the local $SFE$ and the surface
brightness (density) of discs at a given $R$ (beyond their central
regions). To account for the observed surface densities of discs
in the frame of a simple conservative model of evolution of gas
content (a toy model) it is necessary for the local parameter $N$
in the Schmidt law for a disc ($SFR \sim \sigma_{gas}^N$) not to
exceed unit. Only in this case it is possible to reconcile the
observed dependencies $\sigma_{gas}(R)$ and $SFE(R)$ assuming a
gas accretion, more intense for the inner regions of galaxies.
\end{abstract}

\maketitle

\section{INTRODUCTION}

The properties of spiral galaxies determining the star-formation
rate (SFR) at present and in the past are essential for our
understanding of their evolution.

Spiral galaxies are characterized by a very wide range of both the
total SFR and the star-formation efficiency, $\textrm{SFE} =
\textrm{SFR}/M_{gas}$ (star-formation rate per unit mass of gas).
Since the active formation of the stellar population started at
the same epoch in most galaxies (about 10~Gyr ago), we expect a
close relation between these parameters and such galactic
parameters as the total or relative mass of gas in the galaxy
disk. Surprisingly, no such close relations are observed. Like the
relative mass of gas in the disk, the SFE varies over more than an
order of magnitude among galaxies of the same morphological type
or the same color index~\cite{Verhetal01,BBPG01}, while the
average surface densities of (both atomic and molecular) gas
$\sigma_{gas}$, display a fairly small scatter for various types
of spiral galaxies (see, e.g.,~\cite{ZasSmir05} and references
therein). Galaxies containing large masses of gas can be found
among spiral galaxies of all morphological subtypes, although with
different occurrence rates. This reflects the complex nature of
the evolution of galactic disks, which precludes a unique
dependence of the SFR on the morphological features of a galaxy or
the amount of interstellar gas it contains.

Nevertheless, observations of relatively nearby galaxies show the
local gas density to be a crucial factor in star formation.
Indeed, regions of active star formation are always associated
with enhanced gas densities (on scale lengths of several hundred
parsecs and more), whereas star formation is virtually absent in
galaxies with low gas contents (such as lenticular galaxies), or
is restricted in isolated regions where the gas is concentrated.
The relationship between the SFR and the local gas volume density,
$\rho_{gas}$, is often presented in the form proposed by
Schmidt~\cite{Schmidt59} for our own Galaxy: $\textrm{SFR}
\sim\rho_{gas}^n$, where $n\approx2$. In other galaxies, we
directly measure the surface (and not volume) gas density,
$\sigma_{gas}$, which we compare to the ``surface'' SFR, and the
above relation is usually written in the form $\textrm{SFR}
\sim\sigma_{gas}^N$ (we refer to this as the Schmidt law for the
disk, since it is by no means the same as the ``volume'' Schmidt
law; see, e.g., the discussion of this issue by
Tutukov~\cite{Tutukov06}).

A constant SFR per unit mass of gas corresponds to $N=n=1$.
However, estimates for local regions may differ from those for
entire galaxies, since the SFR depends not only on the gas
density, but also on other parameters that vary with distance from
the disk center or from galaxy to galaxy. Therefore, parameter $N$
in the integrated Schmidt law reflects the effect of a great
variety of factors that affect star formation, and it is
essentially purely empirical.

According to the estimates of Kennicutt~\cite{Kennicutt98}, this
parameter is $N\approx1.40\pm$0.15 for the integrated gas masses
and SFRs in galaxies derived from the H$\alpha$ line intensity.
SFR estimates inferred from UV observations near 2000~\AA\ yield
$N=1.0{-}1.6$~\cite{Buatetal89} (the large range of possible
values is due to uncertainties in the amount of UV absorption).
However, these estimates were obtained by comparing integrated
quantities for galaxies with obviously different star-formation
conditions. Comparisons of local (azimuthally averaged) SFRs and
gas densities $\sigma_{gas}$ along the galactocentric radii of
particular spiral galaxies are of considerable interest. In their
analysis for several galaxies, Wong and Blitz~\cite{WB02} found
this parameter to be $N=1.1$ or $N=1.7$, depending on the method
employed to calculate the H$\alpha$ absorption. Heyer et
al.~\cite{HCSY04} found the record high value of $N\approx3.3$ for
M33. However, $N$ is close to unity if we associate the SFR
exclusively with the dense molecular component of the interstellar
gas in galaxies~\cite{WB02,GC04,KS05}.

It is evident that the coefficient of proportionality between the
SFR and $\sigma _{gas}^N$ is not constant, and may, in turn,
depend on other factors that vary with $R$ or time. In simulations
of the evolution of star formation, these factors are usually
taken into account either by including ``young stars--gas''
feedback, which regulates the SFR~~\cite{Tutukov06,FT94}, or by
introducing simple analytical relations between the SFR and other
local parameters that depend on the radial coordinate $R$. Various
authors have adopted for such parameters the surface density of
the stellar and gaseous disks, angular velocity of the disk,
velocity of gas entering the spiral arms, which depends on the
difference of the tangential velocities of the disk rotation and
the spiral pattern, the ratio of the gas density to the critical
density for gravitational instability, and the rate of accretion
of gas onto the galactic disk
(see~\cite{Kennicutt98,WB02,BP00,MK01,Bois03} and references in
these papers). Our understanding of which factors actually play
the most crucial roles in star formation is still poor.

Another unsolved problem related to the star-formation history is
associated with the ages of the disks, or the time elapsed since
the epoch of most intense star formation, which may be several
billion years less than the ages of spherical stellar components.
The comparison of the dependence between the central surface
brightness of the disk and the relative mass of gas with models
for the evolution of galaxy disks with monotonically varying SFRs
led McGaugh and de~Blok~\cite{MB97} to conclude that the range of
ages of galactic disks may reach several billion years. Boissier
et al.~\cite{BBPG01} concluded that low-mass galaxies have lower
ages, since massive galaxies have lower relative amounts of gas
and higher heavy-element abundances for the same SFEs. However,
this conclusion was based on a simple model using the integrated
parameters of galaxies averaged over the entire disk, and the
fairly arbitrary assumption that the SFR is proportional to the
product of $\sigma_{gas}^{1.5}$ and the angular velocity of
rotation $\Omega$. It would be more correct here to consider that
we are dealing with differences in the star-formation histories in
the disks, rather than difference in the disk ages. Current SFRs
combined with broadband color indices are quite consistent with
the hypothesis that most galaxies of various luminosities have
similar ages (${\sim}$10--15~Gyr; see,
e.g.,~\cite{KNC94,buzzoni05}).

Comparisons of the amount of observed gas in galaxies, SFRs, and
the surface densities of stellar disks that have formed from the
gas provide important information that can be used to associate
the current SFE with the average SFE for the entire period of star
formation, and estimate the possible contribution of accretion to
the evolution of the gaseous medium. To this end, it is expedient
to analyze the distribution of gas and the star-formation
intensity in the disks of well-studied galaxies at various
galactocentric distances.

In this paper, we analyze the four relatively nearby spiral
galaxies with fairly active star formation: M33, M81, M100, and
M101. Our adopted parameters for these galaxies are listed in
Table~1, whose columns give (1) the name of the galaxy; (2) the
adopted distance to the galaxy, $D$; (3) the disk inclination,
$i$; (4) the semi-major axis of the $25^m$/arcsec$^2$ $B$
isophote; (5) the maximum galactocentric distance $R_{lim}$ for
which data are available for estimating the SFR; and (6) the
morphological type of the galaxy. The galaxies considered do not
have massive companions (although the shape of the spiral pattern
of M101 may bear signs of the interaction of this galaxy with the
surroundings) and have large angular sizes, enabling the use of
IRAS brightness profiles despite the low resolution of the
100~$mu$m IRAS profile. It is also known the azimuthally averaged
distributions of the atomic and molecular gas, the brightness
distributions in various wavelength intervals, and the variations
of the rotational velocity with galactocentric radius for all four
galaxies.

\begin{table}[t!]
\caption{Parameters of galaxies}
\begin{tabular}{l|c|c|c|c|l}
\hline \multicolumn{1}{c|}{Galaxy}&\multicolumn{1}{c|}{$D$,
Mpc}&${i}$&\multicolumn{1}{c|}{$D_{25}/2$}&
\multicolumn{1}{c|}{$R_{lim}$}&\multicolumn{1}{c}{Type}\\
\hline \multicolumn{1}{c|}{1}&\multicolumn{1}{c|}{2} &3
&\multicolumn{1}{c|}{4} &
\multicolumn{1}{c|}{5}&\multicolumn{1}{c}{6}\\
\hline
M33&0.70&$\phantom{^\circ}55^{\circ}$&\multicolumn{1}{c|}{$35.4'$}&
\multicolumn{1}{c|}{$32'\phantom{1}$}&SAcd\\
M81&3.63&59&13.45&11 &SAab\\
M100&17.00&27&3.7&4.2&SABbc\\
M101&7.48&21&14.4&8.4&SABcd\\
\hline
\end{tabular}
\end{table}

The initial data for the galactic disks are listed in Table~2,
whose columns give (1) the number of the galaxy; (2) and (4) the
color index and central brightness used to calculate the mass of
the disk; (3) the radial scale length of the brightness of the
photometric disk; (5) the velocity of the spiral pattern for the
adopted distances and inclination angles from Table~1; and (6)
references for the galactic rotation curves employed.

\begin{table*}[t!]
\caption{Initial data for galactic disks}
\begin{tabular}{l|c|c|l|c|c}
\hline \multicolumn{1}{c|}{Galaxy}&Color index&Brightness scale
length& \multicolumn{1}{c|}{Central
brightness}&$\Omega_p$,~km~s$^{-1}$~kpc&
Rotation curves\\
\hline
\multicolumn{1}{c|}{1}&2&3&\multicolumn{1}{c|}{4}&5&6\\
\hline
M33&$V-I=1.00$~\cite{Lauer}&$5.8'$~\cite{RegVog}&${(m_K)}_0=17.8^m$~\cite{RegVog}&17.7\,\cite{Newton}&\cite{Corb}\\
M81&$B-V=0.81$~\cite{Leda}&$158''$~\cite{Baggett}&${(m_V)}_0=19.9$~\cite{Baggett}&24.5\,\cite{Westpfahl}&\cite{STHTTKT99}\\
M100&$V-I=1.09$~\cite{Beckman}&$48.5''$~\cite{de Jong}&${(m_K)}_0=17.3$~\cite{de Jong}&22.5\,\cite{Hernandez}&\cite{STHTTKT99}\\
M101&$B-V=0.44$~\cite{Leda}&$128''$~\cite{Knapen}&${(m_K)}_0=17.5$~\cite{Knapen}&17.4\,\cite{Elmegreen}&\cite{STHTTKT99}\\
\hline
\end{tabular}
\end{table*}

Below we analyze the dependence of the SFRs on gas density and the
kinematic parameters of the gaseous disk, in order to identify the
conditions under which the mass of the disk, observed SFRs, and
local values of the azimuthally averaged gas density can be
compatible with each other for about the same (cosmological) age
of the galaxies\footnote{Radial SFR profiles for the galaxies
considered can also be found in earlier papers (see,
e.g.,~\cite{HCSY04} for M33,~\cite{Bois03,Buat89} for
M81,~\cite{WB02,Bois03} for M100, and~\cite{WB02} for M101). The
SFRs in these papers were estimated from data in a single
wavelength interval: either H$\alpha$, UV, or far-IR data.}.

\section{ESTIMATION OF THE STAR-FORMATION RATES}

Various methods have been used to estimate the SFR, all based on
analyzing observational manifestations of young stars and
comparing observational data with models for the stellar
population (for a comparison of various methods, see,
e.g.,~\cite{Kenn98,IgP04}. The most commonly used methods are
those where the initial data are either surface brightnesses or
luminosities of galaxies measured in emission lines (usually
H$\alpha$), or the nonionizing part of the UV spectrum, or in the
far infrared: FIR ($\sim$40--120~$\mu$m) or TIR
($\sim$3--1100~$\mu$m), since short-lived young stars in galaxies
with ongoing star formation contribute most of their radiative
energy in these spectral intervals\footnote{We performed the FIR
to TIR translation using the bolometric correction adopted
in~\cite{DHCSK01}.}.

The galaxy brightnesses in emission lines, the UV, and the far IR
used as diagnostics of the SFR should not yield identical results,
because the main contributions to the luminosities in these
spectral intervals are provided by young stars of different ages
(several million years for emission-line regions, several tens of
million years for UV data, and ${\sim} 10^8$~yr for the far IR).
However, this should not be important in the case of a smoothly
varying SFR averaged over a sufficiently large region of the disk.
The fact that all these methods are to some extent model dependent
is a more important source of discrepancies in the results.
However, the most serious problem is taking into account
extinction at optical and UV wavelengths (especially given the
concentration of the absorbing medium in star-formation regions)
and the contribution of old stars to the TIR radiation. The
extinction in H$\alpha$ is usually determined from the relative
intensities of the hydrogen lines (the Balmer decrement), but such
estimates can be inaccurate, since they depend on both the optical
properties of the absorbing dust and its distribution along the
line of sight. Moreover, we have no prior knowledge of the
fractions of the ionizing radiation that are absorbed by dust or
escape from the disk, thereby not participating in the formation
of lines. As a consequence, systematic differences can arise
between SFR estimates obtained using different methods, as has
been pointed out repeatedly by various
authors~\cite{STEBMD00,TM98,GBELC99}. Note also that SFR estimates
based on emission-line intensities are very sensitive to the
adopted form of the initial mass function (IMF) in the domain of
the most massive gas-ionizing stars.

Allowing for extinction and the clumpiness of the brightness
distribution is not likely to be a problem for the FIR (or TIR)
radiation, however the radiation of cool dust whose heating is not
due to young stars is difficult to take into account. The
contribution of this radiation is small for galaxies with high or
moderate star-formation rate, but can be important in galaxies
containing almost no young stars. Finally, the accuracy with which
of SFRs can be compared with the current amount of gas is limited
by our (poor) knowledge of the IMF for low-mass stars, which
provide the main contribution to the mass of the stellar disk.
However, this is true of all methods used to estimate SFRs.

Nevertheless, correcting for extinction based on the FIR/UV ratio,
independent estimates of the SFR using H$\alpha$ and UV emission
data yield results that are consistent to within a factor of three
for galaxies with luminosities $L_{FIR}$ values differing by
several orders of magnitude~\cite{BBGB02}.

Following~\cite{main,BBGB02,HBI03}, we preferred for the galaxies
considered here to use SFR estimates obtained from both UV and
far-IR brightness measurements. This combined method assumes that
all the energy absorbed in the UV is radiated in the far IR, so
that the the UV/FIR ratio provides a quantitative criterion for
this absorption. The results depend only slightly on the optical
properties of dust or the spatial distributions of the dust and
stars, making it stable with regard to details of the model
(see~\cite{IgP04} for a discussion of this issue). We used
smoothed (up to ${\sim} 100''$) UV brightness profiles and UV
absorption estimates derived from the observed  FIR/UV  brightness
ratios taken from Boissier et al.~\cite{main}, who adopted UV
profiles based on measurements with a stratospheric telescope
(FOCA). We adopted the radial gas-surface density profiles
of~\cite{Corb} for M33,~\cite{Rots} (HI) and~\cite{main} (H$_2$)
for M81,~\cite{main} for M100, and~\cite{WB02} for M101. Boissier
et al.~\cite{main} determined the mass of molecular gas including
the dependence of the conversion factor on the heavy-element
abundance $Z$.

Following~\cite{BBGB02,HBI03}, we write the SFR in the form
\begin{equation}\label{5}
{\textrm{SFR}}=\frac{{\textrm{SFR(UV)}}}{1-\varepsilon} .
\end{equation}
Here, $1-\varepsilon$ is the fraction of UV photons that are not
absorbed, based on the  FIR/UV ratio,
${\textrm{SFR(UV)}}=C_{2000}L_{2000}$ is the SFR derived from the
intensity of UV radiation (near 2000~\AA) in the absence of
absorption, $C_{2000}$ the coefficient of proportionality
determined from the stellar-population model, and
$L_{2000}$~[erg~s$^{-1}$\AA$^{-1}$pc$^{-2}$] the monochromatic
luminosity in the 2000\,\AA~line. The coefficient $C_{2000}$ is
equal to $2.03\times10^{-40}$
($M_{\odot}$/yr)/(erg~s$^{-1}$\AA$^{-1}$)~\cite{BBGB02} for a
model stellar population with a normal heavy-element abundance $z
= z_{\odot}$ and Salpeter IMF in the interval
0.1--100~$M_{\odot}$.

The resulting SFR estimates depend on the adopted IMF. One direct
way to refine the IMF is to compare the observed and model
mass-to-luminosity ratios for the stellar population. Comparing
this ratio to the broadband color indices of galaxies, Bell and de
Jong~\cite{BdeJ01} concluded that the Salpeter IMF overestimates
the mass by at least 30$\%$ due to overestimation of the number of
low-mass stars. We take this into account when comparing the
observed and calculated disk densities below.

\begin{figure}[t!]
\includegraphics[scale=0.95]{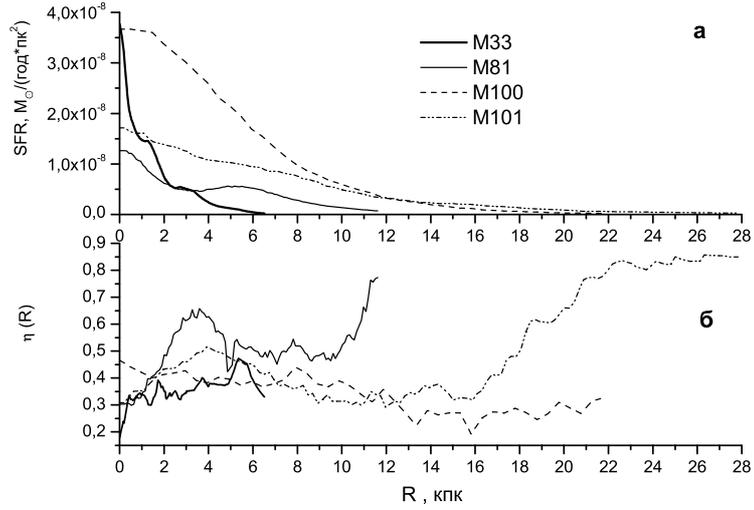}
\caption{Dependence of the (a) SFR [Solar masses /year/pc$^2$] and
(b) contribution of old stars to the IR radiation (TIR),
$\eta(R)$, on the galactocentric distance R [kpc]. \hfill}
\end{figure}

Figure~1a shows radial profiles of the SFR per unit disk area. In
all the galaxies, the SFR decreases with galactocentric distance.
M33 and M100 exhibit the sharpest and smoothest decreases,
respectively, consistent with the different radial scale lengths
of their disks (Table~2). The star-formation intensity is highest
in the central regions of M33 and M101 and the lowest in those of
M81 and M100. Column~5 of Table~3 lists the SFRs integrated over
the disk areas corresponding to $R_{lim}$~(Table~1). (The other
columns of Table~3 list the (1) galaxy number, (2) central density
of the exponential disk estimated by modeling the rotation curve,
and (3)--(4) integrated accretion rates inferred for the
photometric and kinematic disk masses, respectively.)

\begin{table}[b!]
\caption{Integrated accretion rates and the present-time SFRs}
\begin{center}
\begin{tabular}{l|c|c|c|c}
\hline \multicolumn{1}{c|}{Galaxy}&
\multicolumn{1}{c|}{\parbox[c][1cm]{1.3cm}{${S}_0$ ${
M}_{\odot}/\textrm{pc}^2$}}&
\parbox[c][1cm]{1.3cm}{${\textrm{Accr}}_{phot}$ ${M}_{\odot}/\textrm{yr}$}&
\parbox[c][1cm]{1.3cm}{${\textrm{Accr}}_{kin}$ ${ M}_{\odot}/\textrm{yr}$}&
\parbox[c][1cm]{1.3cm}{SFR ${ M}_{\odot}/\textrm{yr}$}\\
\hline
\multicolumn{1}{c|}{1}&\multicolumn{1}{c|}{2}&3&4&5\\
\hline
M33&\multicolumn{1}{c|}{\phantom{11}$636$~\cite{Corb}}&0.3&0.3&0.4\\
M81&660$^\star$&1.1&0.9&1.3\\
M100&1070$^\star$&5.3&5.0&6.9\\
M101&710$^\star$&3.6&\phantom{1}3.54&4.6\\
\hline
\end{tabular}
\end{center}
\footnotesize{$^\star$~Estimate obtained in this paper.}
\end{table}

Our comparison of the SFR profiles with profiles calculated from
the H$\alpha$-line brightness distribution~\cite{MK01} using the
relation~\cite{HBI03}
\begin{equation}\label{101}
{\textrm{SFR}}_{{\textrm{H}}\alpha}
[M_{\odot}/\textrm{yr}]=\frac{L_{{\textrm{H}}\alpha}^{corr}
[\textrm{erg}/\textrm{s}]}{1.25\times10^{-41}}
\end{equation}
and allowing for the line absorption as a function of $R$ in
accordance with~\cite{main} showed a satisfactory agreement (to
within a factor of 1.5) between the two profiles, except for the
1--2~kpc central region and peripheral disk regions, where this
method (UV~$+$~FIR) always yields higher estimates.

Part of the far-IR flux is due not to young stars, but to older
stars whose ages exceed  $10^8$~yr. Like Hirashita et
al.~\cite{HBI03}, we use the radial profile of the fraction of the
UV brightness absorbed, $\varepsilon(R)$, to calculate the factor
$\eta$, equal to the fraction of the TIR flux due to heating of
dust by old stars of the disk (Fig.~1b). As expected, the
contribution of old stars is minimal in the galaxies with the
highest star-formation intensities (M33 and M100). The factor
$\eta$ lies in the range 0.3---0.7 for most regions in all four
galaxies. For comparison, note that the average $\eta$ value
estimated by Hirashita et al.~\cite{HBI03} for galaxies with
moderate SFRs is 0.40$\pm0.06$.

\section{STAR-FORMATION EFFICIENCY: RELATION TO OTHER PARAMETERS}

The star-formation efficiency SFE (SFR per unit mass of gas) is
determined by the relation
\begin{equation}\label{12}
{\textrm{SFE}}
[\textrm{yr}^{-1}]=\frac{{\textrm{SFR}}}{1.4\cdot\sigma_{HI+H_2}}~,
\end{equation}
where the factor of 1.4 accounts for elements heavier than
hydrogen.

\begin{figure}[t!]
\includegraphics{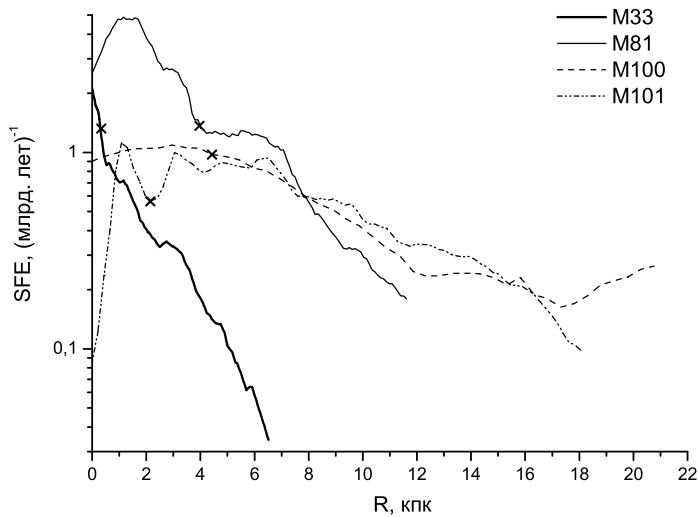}
\caption{Dependence of the SFE [($10^9$ years)$^{-1}$] on the
galactocentric distance R [kpc]. The crosses indicate the adopted
boundaries of inner regions. \hfill}
\end{figure}

The inverse of the SFE corresponds to the time scale for  gas
depletion (without allowance for gas returned to the interstellar
medium)\footnote{This process can lengthen the gas-depletion time
scale by a factor of two to three for galaxies with low fractions
of remaining gas~\cite{KNC94}.}:
\begin{equation}\label{13}
\tau_{gas}, \textrm{Gyr}=\frac{1}{10^9\cdot {\textrm{SFE}}}
\end{equation}
Figure~2 shows the distribution of SFE$(R)$. The crosses indicate
the adopted boundaries for the inner regions of galaxies, where
the estimates are more uncertain. We define these inner (central)
regions to have no clear spiral pattern and/or a size smaller than
the smoothing scale, equal to  $1.5'$~\cite{RegVog},
$3.8'$~\cite{Westpfahl,Schweizer}, $45''$~\cite{Hernandez}, and
$1'$~\cite{Kenney} for M33, M81, M100, and M101, respectively.

The SFR decreases with galactocentric distance throughout most of
the disks in all the galaxies except M101. In this last galaxy,
the monotonic decrease of the SFR begins only at
$R\approx6{-}7$~kpc, where the brightest and most extended HII
regions in the spiral arms are observed. M33 has the lowest SFE
(except for its central region).

The SFR reflects the efficiency of various factors stimulating
ongoing star formation. In generally accepted picture of star
formation in galaxy disks, the following factors should be most
important for the enhancement of this process:

-- a high surface gas density at a given $R$, which also means the
opacity to ionizing radiation and UV radiation that destroys
molecules;

-- a high angular velocity of rotation of the disk~\cite{HEB98};

-- high velocity of gas relative to the spiral density wave that
leads to gas compessing(far from the corotation);

-- a high ratio of the surface gas density to the critical surface
density for gravitational instability of the rotating gaseous
layer;

-- a high density of the stellar disk, which determines (for a
given dispersion of velocities of the gas) the thickness and,
consequently, the volume density of the gaseous disk, and also the
density of the interstellar medium (see, e.g.,~\cite{Elm87}).

We now use these four galaxies to analyze the dependence of the
SFE on the factors counted above.

\subsection{Dependence of SFE on Gas Surface Density}

\begin{figure}[t!]
\includegraphics{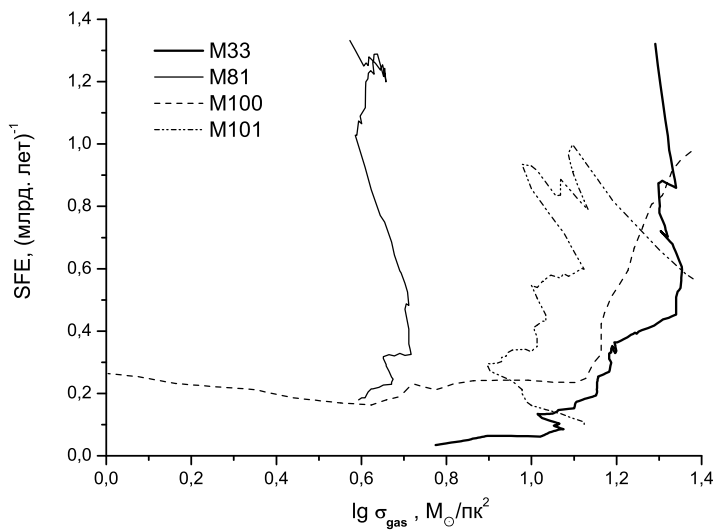}
\caption{Dependence of the SFE [($10^9$ years)$^{-1}$] on the
logarithm of the gas surface density [Solar mass/pc$^2$]. \hfill}
\end{figure}

It follows from Fig.~3 that there is no single dependence between
theese parameters for all four galaxies and, moreover on the
whole, the quantities considered exhibit no clear correlations.
However, M100 and M33 show an essentially monotonic increase of
the SFE with the gas density over a wide range of $\sigma_{gas}$;
i.e., the exponent of the Schmidt law for the disk exceeds unity,
$N>1$\footnote{The increase of the SFE with $\sigma_{gas}$ in M33
can be seen only after smoothing of the curve. The almost vertical
portion on the plot corresponds to the inner region of the galaxy
with a radius of ${\sim} 0.6{-}1$~kpc.}.

The overall picture remains unchanged if we estimate the SFE using
only the mass of molecular gas instead of the total mass of gas:
all four galaxies show different behavior. Note that only one of
the four galaxies (M100) contains a large amount of molecular gas.
The SFR$/M_{{\textrm{H}}_2}$ ratio in this galaxy systematically
decreases with increasing surface density of H$_2$, from the
peripheral regions of the galaxy to the central molecular disk,
which has a very high gas surface density. A similar conclusion
about the decrease of the integrated value of this ratio with
increasing fractional mass of molecular gas was obtained earlier
in~\cite{KasZas06}.

\subsection{Dependence of the SFE on the Angular Rotational Velocity}

\begin{figure}[t!]
\includegraphics{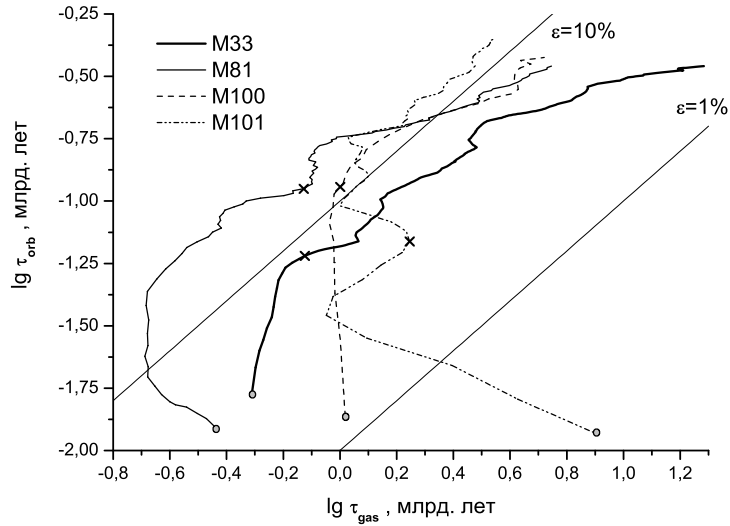}
\caption{Logarithmic dependence of the orbital rotational period,
${\tau}_{orb}$ [$10^9$ years], on the gas-depletion time scale,
${\tau}_{gas}= {\textrm{SFE}}^{-1}$[$10^9$ years]. The circles
indicate the galaxy centers and the crosses the adopted boundaries
of the inner regions. \hfill}
\end{figure}

Arguments in the favor of such a relation can be found in the
papers by Hunter et al.~\cite{HEB98} and
Kennicutt~\cite{Kennicutt98}. According to the latter estimates,
about 10$\%$ of the total gas mass is converted into stars during
one revolution of the galaxy. For the sake of illustration, Fig.~4
compares quantities having the same dimensions: the rotational
period, $\tau_{orb} = 2\pi R / V_c$, and the gas-depletion time
scale, $\tau_{gas} = {\textrm{SFE}}^{-1}$. The gas-depletion time
scales differ widely in different galaxies with the same
rotational period. M33, M81, and M100 seem to show a monotonic
increase in the gas-depletion time scale with increasing
rotational period beyond their central regions (indicated by the
crosses in the figures). However, this increase is nonlinear and
nonmonotonic. The relationship between these two quantities in the
inner region of M101 (${\tau}_{orb}<3\times10^7$~yr) shows instead
the opposite behavior. On the whole, about 1$\%$--10$\%$ of the
gas is converted into stars during each revolution of the galaxy
at these galactocentric distances. The gas depletion is most rapid
in M81 and M101, amounting to 15$\%$ of the gas per revolution at
the galaxy periphery. Wong and Blitz~\cite{WB02}, who, unlike us,
estimated the SFRs from H$\alpha$ brightness measurements, drew
similar conclusions.

\subsection{Dependence of the SFE on the Relative Linear Velocity of
the Disk and the Spiral Pattern}

\begin{figure*}[t!]
\includegraphics{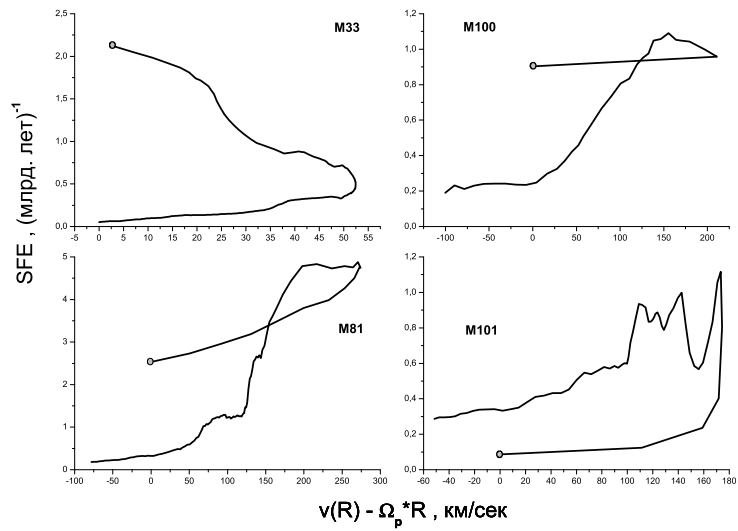}
\caption{Dependence of the SFE [($10^9$ years)$^{-1}$] on the
relative velocity of the disk and spiral pattern [km/s] at a given
radius $R$. The circles indicate the galaxy centers. \hfill}
\end{figure*}

We determined this relative velocity as $\Delta V=|V_c(R)-
\Omega_pR|$, where $\Omega_p$ is the pattern speed of the density
wave, so $\Delta V$ vanishes at the corotation radius $R_C$, where
$V_c(R_C) = \Omega_pR_C$. The pattern speeds $\Omega_p$ for the
four galaxies were measured by various authors; the values used
below and the corresponding references can be found in Table~2. If
the relative velocity were the key factor for star formation, the
minimum SFE would be located at the corotation radius, and the SFE
would increase on either side of the corotation radius, passing
through a maximum in the inner region at $R\approx R_C/2$, where
the relative velocity of the pattern and the disk is maximum. In
three of the four galaxies (M81, M100, and M101), the measurements
cover regions beyond $R_C$, and the SFE$(\Delta V)$ dependence
does not fit the expectations (Fig.~5). This clearly demonstrates
that the density wave imposes order on rather than triggers star
formation: the increase in the SFE in the spiral arms, which can
be important (see, e.g.,~\cite{LY90}), is compensated by its
decrease in inter-arm regions.

The situation in M33 supports this conclusion. In this galaxy, the
SFE increases with distance from the corotation radius to the very
center, despite the fact that $\Delta V$ should decrease in the
central region due to the decrease of the linear rotational
velocities of both the disk and spiral pattern.

\subsection{The Role of Gravitational Instability of the Gaseous Layer}

At a certain surface density, $\sigma_{crit}$, which depends on
the kinematic parameters of the disk, the gaseous layer, becomes
gravitationally unstable on scale lengths of several kpc. The
growth of instability can result in the formation of gas
complexes, with subsequent star formation. We have for a thin
gaseous layer:
\begin{equation}\label{100}
\sigma_{crit}=\frac{2 c_g \Omega}{\pi G Q_T}~ \sqrt{1+\frac{R}{2
\Omega}\frac{d\Omega}{dR}}=\frac{2}{\pi G} \frac{c_g}{Q_T}~
\frac{V}{R}~\sqrt{\frac12+\frac{R}{2V}\frac{dV}{dR}},
\end{equation}
where $c_g\approx6{-}8$~km/s is the one-dimensional velocity
dispersion for the gaseous clouds and $Q_T\ge1$ is the Toomre
dimensionless stability parameter. According to the estimates of
Martin and Kennicutt~\cite{MK01}, the observed sharp decrease of
the SFR at a definite galactocentric distance corresponds to
${c_g}/{Q_T}\approx4$~km/s.

Weak star formation is known to occur even in regions where the
average gas surface density is certainly lower than the critical
surface density (in low surface brightness galaxies, lenticular
galaxies, and at the peripheries of spiral galaxies). Therefore,
the condition $\sigma_{gas}>\sigma_{crit}$ (or approximate
equality of these quantities) is not required for star formation.
A role for gravitational instability is, nevertheless, supported
by the observational data: despite the approximate nature of
$\sigma_{crit}$ estimates, the gas surface densities at different
galactocentric distances in spiral galaxies usually differ from
the critical density by no more than a factor of two in either
sense, and star formation almost ceases in regions where the
gaseous disk is certainly stable, say, at the periphery of the
disk (see~\cite{ZasSmir05,WB02,MK01} for a discussion).

\begin{figure}[b!]
\includegraphics{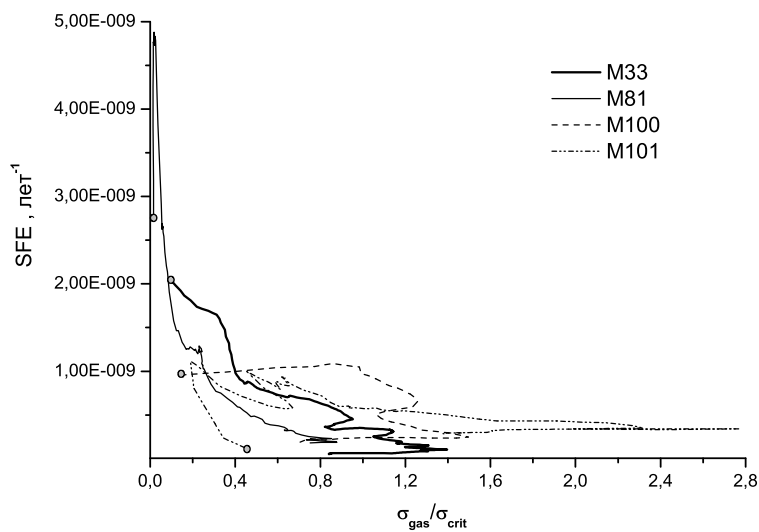}
\caption{Dependence of the SFE [(years)$^{-1}$] on the ratio of
the gas surface density to the critical density
$\sigma_{gas}/\sigma_{crit}$ at a given radius. The circles
indicate the galaxy centers. \hfill}
\end{figure}

Figure~6 shows the dependence of the SFE on the ratio
$\sigma_{gas}/\sigma_{crit}$. Note, however, that regions with
$\sigma_{gas}/\sigma_{crit}<1$ that are far from the center remain
uncovered by the data. None of the four galaxies shows a
systematic increase in the SFE with increasing azimuthally
averaged $\sigma_{gas}/\sigma_{crit}$. We conclude that, in the
range of $\sigma_{gas}/\sigma_{crit}$ considered, this ratio has
only a weak effect on the SFE. Moreover, regions with low
$\sigma_{gas}/\sigma_{crit}$ can be characterized by a high SFE,
but this is due not to intense star formation, but instead due to
low gas density. In this regions the birth of stars is triggered
by other factors that are not associated with large-scale
gravitational instability. However, a quantitative analysis of the
role of instability requires a more refined approach and, in
particular, allowance for the inhomogeneity of the gaseous disk
and the presence of the stellar disk, which also affects the
stability~\cite{Bois03}.

\subsection{Dependence of the SFE on the Azimuthally Averaged
Brightness (Density) of the Stellar Disk}

Figure~7a compares the SFEs and the smoothed surface brightnesses
of the disks, described by a simple exponential dependence on $R$
(in mag/arsec$^2$). Apart from in their central regions, all the
galaxies show a decrease in the SFE with decreasing surface
brightness and, obviously, with decreasing density of the stellar
disk, $\log\sigma_{phot}$. Figure~7b compares the SFE and the
surface density of the disk estimated from the $K$-band brightness
distribution $\mu_K$ and the ratio $M/L_K$, which depends on the
color indices of the galaxies (Table~2) in accordance with the
models for the evolution of the stellar
systems~\cite{BdeJ01}\footnote{We calculated the radial $K$-band
brightness profile for М81 from the radial profile for
$\mu_V(R)$.}.

It is striking that the SFE($\log\sigma_{phot}$) dependence
outside the central regions is virtually the same for all four
galaxies over a wide range of surface densities,
$1.3<\log\sigma_{phot}~[M_{\odot}/\textrm{pc}^2]<2.7$. In other
words, the SFE there is determined by the surface density of the
disk, and is higher the denser the stellar disk\footnote{Earlier,
Bell and de Jong~\cite{BdeJ00} also analyzed the relationship
between the star-formation history and the brightness (density) of
the disk. Sea also Dopita and Ryder~\cite{DR94} and Boisier et
al.~\cite{Bois03} who used the empirical dependence of the SFR on
the product of the surface densities of the gas and disk to
appropriate powers.}.

The outer, more tenuous regions of disks rotate more slowly,
suggesting that the gas-depletion time scale, which is inversely
proportional to the SFE, should increase with the rotational
period, $\tau_{orb}$, as we can see from the observations
(Fig.~4). The straight line in Fig.~7b corresponds to the relation
\begin{equation}\label{303}
\log {\textrm{SFE}} = 0.70\cdot\log\sigma_{phot}-1.73~.
\end{equation}
The average deviation of $\log {\textrm{SFE}}$ from this relation
is about 0.3 in the interval of $\log\sigma_{phot}$ mentioned
above.

We believe that $\sigma_{phot}$ characterizes the surface density
of the stellar disk, $\sigma_{*}$, so that the equation~(6) is
equivalent to the dependence $\textrm{SFR} \sim
\sigma_{gas}\cdot\sigma_{*}^{0.7}$, which reflects the influence
of the disk density on the ongoing star formation.

The decrease of the SFE with decreasing surface density of the
disk, especially in the outer regions of the galaxies, may be
associated with two factors. First, the decrease in the density of
the stellar disk results in an increase in the thickness of the
gaseous layer in it, since the potential well produced by the disk
becomes increasingly ``shallower''. As a consequence, the volume
density and pressure of the gas decrease. The growth of the
thickness of the gaseous layer is most rapid in the outer regions
of the Milky Way, in accordance with the expected balance of
forces determining the equilibrium of this layer~\cite{NJ02}. The
second, fairly obvious, factor is the decrease in the inflow of
gas ejected by evolved stars, which compensates at least partially
for the consumption of gas by star formation, especially if the
surface density of gas is much lower than that of the stellar
disk.

Note that the similarity of the dependences of the SFE and disk
surface brightness is lost completely if we use the density of the
molecular component of the gas instead of the total gas density.

\begin{figure}[t!]
\includegraphics{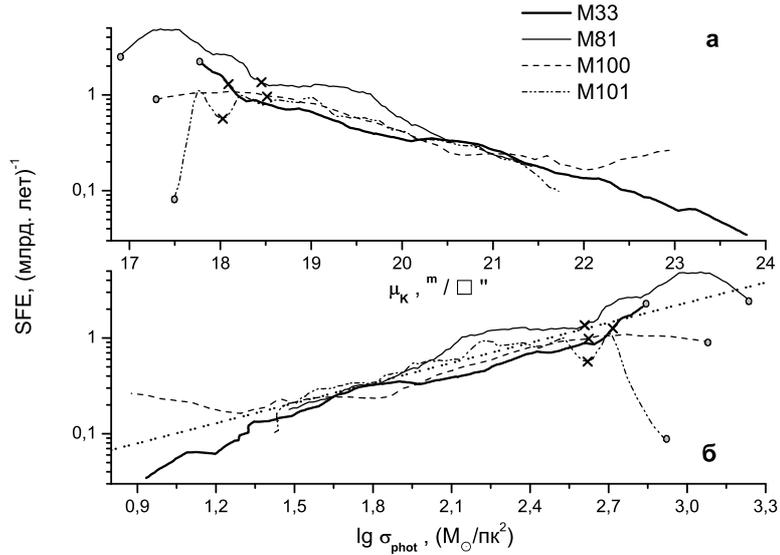}
\caption{Variation of SFE [($10^9$ years)$^{-1}$] as a function of
the (a) surface brightness $\mu_K$ and the (b) logarithm of the
photrometrically derived surface density of the disk,
$\log\sigma_{phot}$[Solar masses/pc$^2$]. The circles indicate the
centers of the galaxies and the crosses the adopted boundaries of
the inner regions. The dashed straight line in (b) corresponds to
the dependence ${\textrm{SFE}}\sim\sigma_{phot}^{0.7}$.  \hfill}
\end{figure}

\section{THE STAR-FORMATION HISTORY\protect\\ IN THE DISK}

If the SFE is primarily determined by the surface density of the
disk, this raises the question of how the observations fit the
simplest model, in which both the total surface density of the
disk and the SFE are time independent and vary only with radius.
The low accuracy of SFE estimates even for well-studied galaxies
(no better than within a factor of two) prevents reliable
estimation of the masses of stars born during past epochs based on
the current amount of gas, but this does not remove the need to
test the consistency of SFE estimates with the kinematically or
photometrically determined disk densities.

\begin{figure*}[t!]
\includegraphics{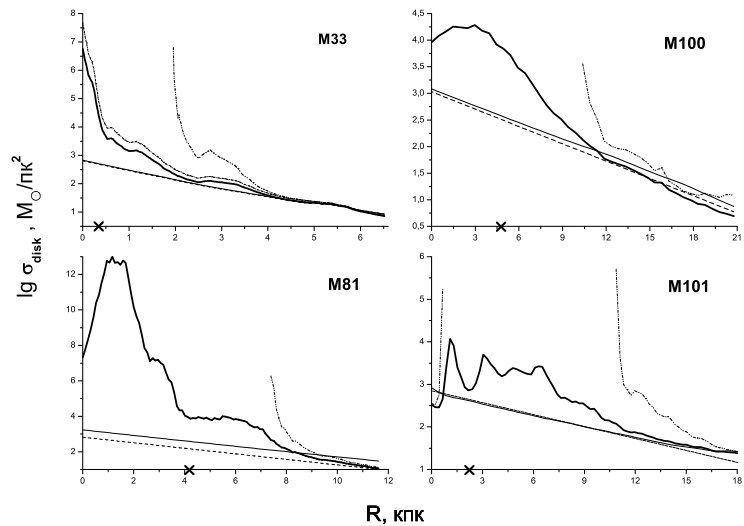}
\caption {The radial distributions of the surface disk densities
$\sigma_{disk}$ [Solar masses / pc$^2$]. The solid bold line shows
the model dependence for $N=1$, $r=0.4$, and the dash
double-dotted line -- the same for $N=1.4$ (a ``toy model''); the
thin solid line and the dashed line show the photometrically and
kinematically determined disk densities (stars plus gas). For
comparison, the dash dotted line shows the profile of the disk
density for $N=1$ and $r=0.3$ (a ``toy model'') for M33. The
crosses on the horizontal axis indicate the boundaries of the
inner regions (see the text). \hfill}
\end{figure*}

We first use an extremely simplified (``toy'') model with no
accretion and no radial motion of gas, with most of the gas
ejected by previous generations of stars being returned to the
interstellar medium before the mass of the remaining gas changes
significantly (the approximation of instanteneous gas return).
This assumption is based on model estimates, which indicate that
most of the gas lost by stars of a single generation returns to
the interstellar medium during the first one to two billion
years~\cite{KNC94,NO06}, but may not be applicable if the fraction
of gas in the disk is too small~\cite{KNC94}.

Note that the gas in galaxy disks usually exhibits no systematic
radial motions (disregarding galaxies with massive bars or
interacting galaxies): measurements made for several nearby
galaxies yielded only lower limits of several km/s for the radial
velocities~\cite{WBB04}.

If we specify $t=0$ to be the time when the disk was entirely
gaseous (the beginning of the evolution) and neglect variations of
the SFE with time, the decrease in the relative gas content at a
given $R$ can be described by the simple law
\begin{equation}\label{102}
\sigma_{gas}/\sigma_{disk}= e^{-{\textrm{SFE}}\cdot(1-r)\cdot t} ,
\end{equation}
where $\sigma_{disk}$ is the total density of the stellar and
gaseous disk and $r$ is the fraction of gas returned by stars to
the interstellar medium. The available estimates usually yield $r
= 0.3-0.4$, depending on the stellar IMF
employed~\cite{KNC94,NO06,JCP01}. At $t = T$, where
$T\approx10^{10}$~yr is the age of the disk (see the
Introduction), the model must explain the current disk density at
a given $R$ starting from the observed density of gas and
estimated SFEs.

In the general case, the total surface density of the disk is
\begin{equation}\label{103}
\sigma_{disk}=\sigma_{gas}\cdot e^{{\textrm{SFE}}\cdot(1-r)\cdot
T},\quad N=1 ,
\end{equation}
\begin{equation}\label{203}
\sigma_{disk}=\frac{\sigma_{gas}}{{\left[1-{\textrm{SFE}} (1-r)
(N-1) T\right]}^{\frac{1}{N-1}}} , \\ \nonumber
 N\ne1 .
\end{equation}

We determined the observed surface densities of the stellar disks
using both modeling of the rotation curve and the photometric
parameters of the disk, as described in the previous section.

For M33 we adopted the density distribution inferred from the
rotation curve taken from~\cite{Corb}, and used the best-fit model
including the stellar and gaseous disks and a dark halo, with the
component masses scaled to the adopted distance (0.7~Mpc). For the
three remaining galaxies, we constructed three-component kinematic
models, each including a King bulge, thin disk with an exponential
density decrease and the radial scale set equal to the photometric
radial scale to within 10$\%$--15$\%$ (Table~2), and a dark halo
with a quasi-isothermal volume-density profile. We determined the
disk parameters from the best fit to the rotation curve, and
although these are not completely free from the effects of
ambiguity in the component masses, a fixing of the radial scale
for the density variations considerably narrows the interval of
possible component masses.

Figure~8 compares the expected disk density, $\sigma_{disk}(R)$,
derived for the simplified model with the given $\sigma_{gas}$(R)
and SFE$(R)$ and with $N=1$, and the density distributions derived
photometrically and kinematically from the observations. Here, we
set the fraction of mass ejected by stars to $r=0.4$, although a
quick check showed that changing this fraction to $r=0.3$ has no
significant effect on the overall picture obtained. As an example,
Fig.~8 shows the radial profile of the disk density for M33
calculated with $r=0.3$. Given the approximate nature of the
estimates, we conclude that, with $N=1$, the estimated disk
densities for the simplified model are consistent with the density
measurements in the outer regions, but yield unacceptably
overestimated densities for the inner parts of the disks. We show
below that this discrepancy disappears in the case of moderate gas
accretion rates. If we, nevertheless, assume that the exponent in
the Schmidt law exceeds unity, $N>1$, then the disk density
extrapolated back to 10~Gyr is too high (Fig.~8, the ``dash
double-dot'' line for each of the galaxies), since in this case
the SFR in the past must be significantly higher than for $N=1$.
Therefore, the condition $N>1$ is incompatible with this simple
model for the evolution, and requires the SFE to be significantly
lower in the past than at the present epoch for the bulk of the
disk stars\footnote{Gallagher et al.\cite{gallagher84} were the
first to deduce the slow variation of the integrated SFR, which is
difficult to reconcile with the Schmidt law.}.

It is easy to verify that, even if we assume that the SFE in the
four galaxies is overestimated by a factor of two, the resulting
estimates for the disk density in the inner regions of the
galaxies remain unrealistically high.

This discrepancy can be removed if we introduce into our model the
accretion of gas onto the disk, which slows the decrease in the
amount of gas with time. Accretion certainly played an important
role in the initial stages of formation of the galactic disks, but
its role in the subsequent evolution is poorly studied. We have
both direct and indirect evidence for ongoing accretion onto the
galactic disks (see, e.g., the discussion of this issue
in~\cite{NO06,BBCPB02}). It is important that even low accretion
rates of ${\sim}1$~$M_{\odot}/$yr can appreciably compensate the
loss of gas to star formation. Of the four galaxies considered,
the effect of accretion is most obvious in the inner region of
M81, where the high SFE (Fig.~2), due to the low gas density,
implies a gas-depletion time scale of only 200--500~Myr if no gas
is fed back to the disk.

\begin{figure*}[t!]
\includegraphics{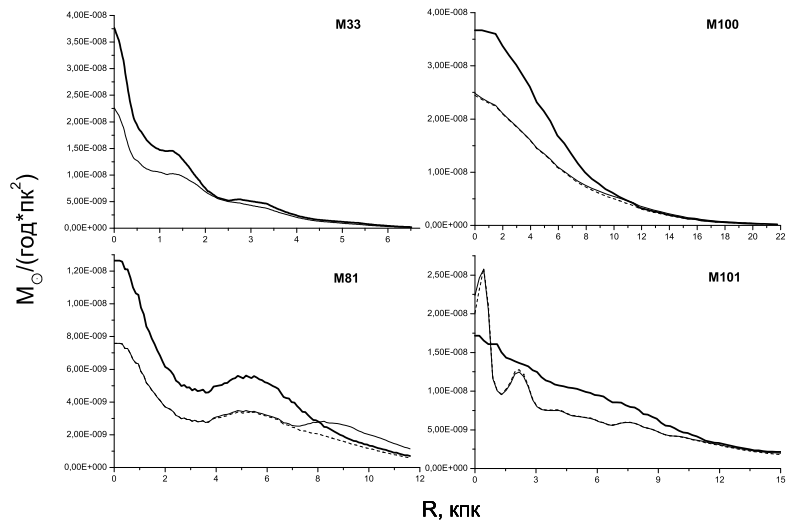}
\caption{Comparison of the SFRs (bold solid line) and the
accretion rates ([Solar masses/year/pc$^2$]) along the radius $R$
[kpc] required to explain the observed disk surface densities. The
thin solid and dashed lines show the results obtained using the
photometrically derived density $\sigma_{phot}$ and the density
$\sigma_{kin}$ derived from the disk kinematics (the photometric
and kinematic densities coincide in M33, so that the two curves
merge). \hfill}
\end{figure*}

The accretion of interagalactic gas plays the same role as the
accretion of gas ejected by long-living stars of the disk or
bulge. The latter accretion enables the gas-depletion time scale
in galaxies with low gas contents to appreciably exceed
$\tau_{gas}={\textrm{SFE}}^{-1}$~\cite{KNC94}. Despite the overall
similarity of the two accretion processes, accretion ``from
outside'', unlike ``internal accretion'', increases the total mass
of the disk and constrains the growth of the heavy-element
abundance in the gas.

Let us now estimate the maximum accretion rates
${\textrm{Accr}}(R)$ for the four galaxies that make the model
disk surface density close to the observational disk density
estimates.

Let the SFR per unit disk area at a given galactocentric distance
be equal to SFR $=A\sigma_{gas}^N$, where $A$ is the coefficient
of proportionality, which, in general, varies with $R$. The
equation relating the SFR, gas density, and accretion rate has the
form
\begin{equation}\label{110}
\frac{d\sigma_{gas}}{dt}={-}(1-r)\cdot
A\cdot\sigma_{gas}^N+{\textrm{Accr}} .
\end{equation}
Here, as above, $r$ is the fraction of gas returned by stars and
${\textrm{Accr}}$ is the accretion rate. The variables in this
equation can be separated:
\begin{equation}\label{111}
\frac{d\sigma_{gas}}{{\textrm{Accr}}-(1-r)\cdot
A\cdot\sigma_{gas}^N}=dt .
\end{equation}
This equation has a simple analytical solution only when $N=1$. In
this case, $A$ is equal to the star-formation efficiency.

The solution for the total mass of stars born has the form
\begin{equation}\label{112}
\frac{1}{(1-r)\cdot {\textrm{SFE}}}\, \ln\left|\frac{(1-r)\cdot
{\textrm{SFE}}\cdot \sigma_{disk}-{\textrm{Accr}}}{(1-r)\cdot
{\textrm{SFE}}\cdot\sigma_{gas}(T)-{\textrm{Accr}}} \right|=T ,
\end{equation}
where $T$ is the age of the disk and the accretion rate is assumed
to be time independent. It follows that
\begin{equation}\label{113}
e^{T\cdot(1-r)\cdot {\textrm{SFE}} }=\left|\frac{(1-r)\cdot
{\textrm{SFE}}\cdot\sigma_{disk}-{\textrm{Accr}}}{(1-r)\cdot
{\textrm{SFE}}\cdot\sigma_{gas}(T)-{\textrm{Accr}}}\right| .
\end{equation}
The expression inside the modulus signs can be either positive or
negative, although the two solutions differ little in the case in
which we are interested, when the model estimate of the disk
density $\sigma_{disk}$ obtained neglecting accretion
significantly exceeds the observed density. The maximum accretion
rate corresponding to negative values of the expression inside the
modulus signs is
\begin{equation}\label{45}
{\textrm{Accr}}_{-}=(1-r)\cdot{\textrm{SFE}}\,
\frac{e^{T\cdot(1-r)\cdot{\textrm{SFE}}}\sigma_{gas}(T)+\sigma_{disk}}{e^{T\cdot(1-r)\cdot{\textrm{SFE}}}+1}
,
\end{equation}

We will now use this equation to estimate for all four galaxies
the accretion rate ${\textrm{Accr}}(R)$ for which the model disk
density at any radius is equal to the observed density derived
from photrometric data or the rotation curve. Figure~9 shows a
comparison of this accretion rate and SFR.

Table~3 lists the corresponding integrated rates of gas accretion
onto the galaxy within $R_{lim}$ (Table.~1). It follows from (9)
that, when the accretion rates are close to the SFRs, the observed
disk densities can be explained by a simple model. Note that the
need for accretion is not obvious for the outer regions of the
galaxies.

The fact that integrated accretion rate and SFR are close to each
other (Table~3) is not just a coincidence: this follows from the
adopted model with $N=1$. The observed disk densities and the
condition $N>1$, or equivalently the increase in the SFR with
increasing gas density, are more difficult to reconcile. In this
case, we must assume that either the SFE was significantly lower
in the past than at present, despite the higher gas density (i.e.,
the condition $N>1$ was not satisfied during the epoch of
formation of the bulk of the disk), or the age of the inner
regions is appreciably lower than the cosmological age. At the
same time, our overall picture of the evolution of the gas content
in the disk becomes self-consistent if we suppose that the
exponent~$N$ in the Schmidt law does not exceed unity at all
galactocentric distances. In this case, the surface densities in
the outer regions of galaxies can be satisfactorily explained
using a simple conservative model. The integral $N>1$ estimates
obtained for the current star formation appear to result from
comparing the SFR and gas density at different galactocentric
distances, where other factors affecting the SFR apart from the
gas density come into play, so that $N$ can no longer be
considered to be local.

To summarize, we emphasize that in none of the galaxies considered
did the Schmidt law, defined as the relation between the
distributions of the SFR and $\sigma_{gas}$, result in good
agreement with the observations. This could be due to the effect
of various factors influencing star formation at different
galactocentric distances. Indeed, the evolution of the gaseous
component should have different behavior in the inner part of the
disk, with radii of several kpc, and in the regions located
farther from the center. The gas content in the outer regions of
the galaxies can be described satisfactorily using a model with a
time-independent SFE for local disk density, but the evolution in
the inner disk regions is more complex that cannot be reproduced
in a conservative model without allowance for the accretion of gas
onto the disk and/or radial motion of the gas. In this case, an
exponential law for the decrease in the SFR proves to be
inapplicable.

Note that a similar situation that requires inclusion of the
accretion of gas also arises during modeling of the evolution of
the disk in our own Galaxy. In the evolutionary model of Naab and
Ostriker~\cite{NO06}, where the SFE was taken to be fixed at a
given galactocentric distance (and equal to 10$\%$ of the disk
rotational period), the ``global'' SFR and accretion rate do not
decrease exponentially with time, but vary slowly and remain close
to each other over the past several billion years. This model
agrees well with available estimates of the surface density of the
stellar disk and gaseous layer in the solar neighborhood.

\section{ACKNOWLEDGMENTS}

We are grateful to A.V.~Tutukov for discussions of this work. This
work was supported by the Russian Foundation for Basic Research
(project code no.~04-02-16518).

\end{document}